\documentclass[pra,reprint,superscriptaddress,amsmath,amssymb,aps,floatfix]{revtex4-1}
\usepackage[utf8]{inputenc}
\usepackage{hyperref}
\usepackage{graphicx}
\usepackage{comment}
\usepackage{braket}

\usepackage{color}
\usepackage[normalem]{ulem} 	
\definecolor{myblue}{rgb}{0.4, 0.3, 0.7}
\newcommand{\asr}[1]{{#1}}
\definecolor{purple}{rgb}{0.63,0,1}

\definecolor{dark-green}{rgb}{0,0.4,0.1}
\definecolor{dark-gray}{rgb}{0.4,0.4,0.4}

\definecolor{pink}{rgb}{1,0,0.9}

\newcommand{\subfigref}[2]{\hyperref[#1]{\ref*{#1}(#2)}}

\begin{document}
\title{
Variational optimization of tensor-network states with the honeycomb-lattice corner transfer matrix
}

\author{I.V. Lukin}
\email{illya.lukin11@gmail.com}
\affiliation{Karazin Kharkiv National University, Svobody Square 4, 61022 Kharkiv, Ukraine}

\author{A.G. Sotnikov}
\email{a\_sotnikov@kipt.kharkov.ua}
\affiliation{Karazin Kharkiv National University, Svobody Square 4, 61022 Kharkiv, Ukraine}
\affiliation{Akhiezer Institute for Theoretical Physics, NSC KIPT, Akademichna 1, 61108 Kharkiv, Ukraine}

\date{\today}

\begin{abstract}
We develop a method of variational optimization of the infinite projected entangled pair states on the honeycomb lattice.  The method is based on the automatic differentiation of the honeycomb-lattice corner transfer matrix renormalization group. We apply the approach to the antiferromagnetic Heisenberg spin-1/2 \asr{and ferromagnetic Kitaev models} on the honeycomb lattice.
The developed formalism gives quantitatively accurate results for the main physical observables and has a necessary potential for further extensions.
\end{abstract}

\maketitle

\maketitle

\section{Introduction} 
The most difficult part of any two- or three-dimensional tensor-network algorithm, especially the infinite projected entangled pair states (iPEPS) optimization and calculation of observables with it, is the tensor-network contraction~\cite{Ran_2020}. Exact contraction of the PEPS norm is exponentially hard in general.
Therefore, in two-dimensional tensor-network calculations, one requires to contract a tensor network approximately. There are three main methods to contract two-dimensional tensor networks with certain translational invariance: transfer matrix methods, also called the boundary matrix product states~\cite{PhysRevLett.101.250602}, tensor renormalization group (TRG)~\cite{PhysRevB.78.205116, PhysRevLett.101.090603} including its modifications, and methods based on the corner transfer matrix (CTM)~\cite{Orus_2009}. 
There are also several mixed versions, e.g., the channel environments ~\cite{Vanderstraeten_2015}, which are similar to both CTM and transfer matrix methods. 

All the mentioned approaches are most naturally defined for a square lattice geometry, though transfer matrices were also used for kagome and honeycomb lattices~\cite{xie2017optimized}, while TRG was applied to the honeycomb lattice~\cite{PhysRevLett.101.090603}. 
At the same time, CTM methods, at least in the iPEPS calculations, were mainly applied to the square lattice. The strategy to deal with other lattices was to map it onto the square lattice and then to employ already developed algorithms. 
This strategy was successfully applied to star-~\cite{Jahromi_2018_star}, ruby-~\cite{Jahromi_2018}, kagome-~\cite{PhysRevB.86.041106}, triangular-~\cite{PhysRevB.85.125116} and honeycomb-lattice geometries~\cite{PhysRevX.2.041013} with the simple or full update as an optimization method for iPEPS. 

Still, the most accurate iPEPS optimization method is the variational update. Originally, it was proposed for the square lattice~\cite{PhysRevB.94.035133, Vanderstraeten_2016}, where the gradients were derived in terms of the CTM tensors or channel environments. 
Later, it was observed that the gradients can be derived with the help of automatic differentiation~\cite{Liao_2019}, originally developed in the machine-learning community. 
In this approach, one needs only to calculate the energy with an approximate contraction scheme by using only differentiable procedures. 
In particular, it can be applied to the lattices other than square with the help of special mappings of these to the square one~\cite{Var_kagome}. 
The variational approach enables studies of the next-nearest-neighbor (NNN) frustrated systems~\cite{hasik2021investigation} and chiral spin liquids~\cite{hasik2022simulating}. 

In this \asr{paper}, we aim to establish whether other efficient contraction methods on other lattices can lead to a stable variational update with gradients derived by the automatic differentiation. 
We focus on a generalization of the corner transfer matrix renormalization group (CTMRG) approach to the honeycomb-lattice tensor network.
We describe the contraction scheme and employ it for the ground-state analysis of the antiferromagnetic Heisenberg model on the honeycomb lattice. 
We employ the \textsc{zygote} autodifferentiation package~\cite{Zygote.jl-2018} to compute gradients and optimize the iPEPS wave function by using the Broyden-Fletcher-Goldfarb-Shanno (BFGS) algorithm.

It should be noted that CTMRG on lattices other than square, in particular, on hyperbolic lattices~\cite{Dani_ka_2015, Dani_ka_2016}, were used in the variational calculations with the interaction-round-a-face (IRF) type tensor network as a variational ansatz~\cite{nishio2004tensor}. 
These optimizations were conducted by using the Nelder-Mead method.

\section{Method and results}
\subsection{Honeycomb CTMRG} 
CTM was originally developed by Baxter as a method of exact solution of certain integrable models on the square lattice~\cite{Baxter1968, Baxter1978}. Later, it was generalized to a triangular lattice~\cite{Tsang1977,Baxter_1980}. Nishino and Okunishi developed CTM into the numerical renormalization group method (CTMRG)~\cite{CTMRG}, which was further generalized to the classical statistical mechanics systems on triangular and hyperbolic lattices~\cite{Ueda_2007, HyperbolicCTM2, HyperbolicCTM3, HyperbolicCTM4, HyperbolicCTM5}. Still, their scheme on the triangular lattice was tailored for the IRF-like models, while we are interested in the vertex like model, since the PEPS norm can be represented as a contraction of the vertex model  partition function. For this purpose, we dualize the construction from Ref.~\cite{HyperbolicCTM4} by obtaining CTMRG for a honeycomb-lattice tensor network from the CTMRG of the IRF classical model on the triangular lattice.

\begin{figure}[t]
  \includegraphics[width=\linewidth]{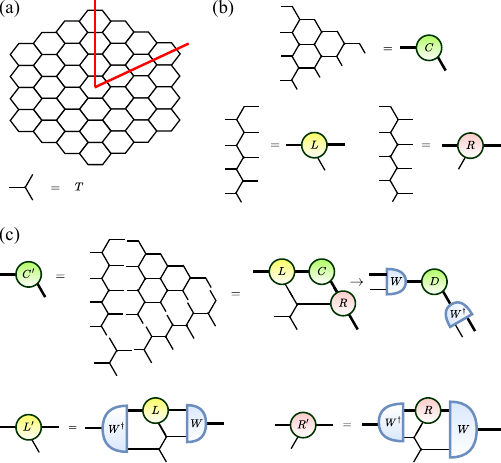}  
  \caption{\label{fig:CTMRG}%
    Illustration of the definitions of CTMRG on a honeycomb lattice. 
    (a) A honeycomb lattice can be divided into six corners. Any trivalent vertex of the lattice corresponds to the tensor $T$. 
    (b) Definitions of corner matrix $C$ and two column tensors $L,R$. In the real calculations we assume that these corners and columns contain an infinite number of sites, but the finite equivalents, shown in the figure, can be used as initialization. (Note that the initialization of the matrix $C$ is Hermitian but not diagonal. One can gauge it to the diagonal form, but since we do not introduce any truncations in the initialization, this additional gauge choice has no impact on any further calculations.)
    (c) Illustration of the update procedure for the matrix $C$ and the tensors $L$ and $R$. This update is repeated until the convergence. 
    }
\end{figure}
Let us start with a honeycomb lattice characterized by a single trivalent tensor~$T$, symmetric under rotations and reflection conjugations, with the bond dimension $D$ on all of its vertices.
This lattice can be divided into six corners, which are shown in Fig.~\subfigref{fig:CTMRG}{a}. Contraction of all tensors in the corner can be represented as a matrix~$C$ [see Fig.~\subfigref{fig:CTMRG}{b}]. If the individual tensors are  symmetric under rotations and homogeneous around the lattice, then all the corner matrices $C$ are identical and the tensor network contraction $Z$ can be represented as
\begin{equation}
    Z = \operatorname{Tr}{C^{6}}.
\end{equation}

Additionally, following Ref.~\cite{HyperbolicCTM4}, we can introduce two row tensors $L$ and $R$, which are also shown in Fig.~\subfigref{fig:CTMRG}{b}. In the infinite system, the tensors $C$, $L$, and $R$ are infinite-dimensional, but in the numerical procedure we can truncate them to a finite dimension by using the spectrum of $C$ as a guide for truncation. 
We initialize $C$, $L$, and $R$ with their small lattice analogs and  increase the lattice by adding lattice sites. In each step of this increase we perform the updates, which are illustrated in Fig.~\subfigref{fig:CTMRG}{c} and can be expressed as  
\begin{align}\label{CTMUpdate}
    C &\to L C R T^{2}, \\
    L &\to L T^{2}, \\
    R &\to R T^{2}.
\end{align}

The updated matrix $C$ has a dimension $\chi D$, where $\chi$ is a dimension of the original matrix $C$. 
Next, we need to truncate the matrix $C$ back to the original dimension~$\chi$. To this end, we diagonalize the matrix $C = W D W^{\dagger}$ and truncate it by using its eigenspectum (note that here the eigenvalues can be negative and must be sorted by the absolute value). 
The Hermiticity of the matrix $C$ is ensured by the reflection-conjugation symmetry of the original tensors $T$. Alternatively, we can use the singular value decomposition (SVD). 
After this decomposition, the new matrix $C$ is set as $C=D$, while the matrices $W$ and $W^{\dagger}$ are absorbed into the updated $L$ and $R$ [see also Fig.~\subfigref{fig:CTMRG}{c}]. 
Finally, we repeat the lattice increasing process until the convergence. The update of the CTM tensors includes only $\chi$ largest eigenvalues of the decomposition, which can be computed with iterative eigensolvers (or a randomized SVD in case of the SVD decomposition) to reduce the complexity. In the case of the randomized SVD decomposition, the computational complexity scales as $O(D^{4}\chi^{3}+D^{6}\chi^{2})$.

\begin{figure}[t]
  \includegraphics[width=\linewidth]{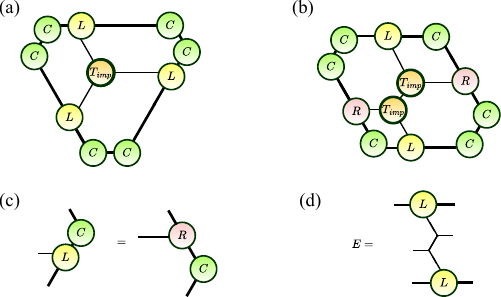}
  \caption{\label{fig:Observables}%
    Illustration of computation of different observables with the CTMRG environments on the honeycomb lattice.
    (a) Calculation of the one-site correlator with the impurity tensor~$T_{\rm imp}$, which characterizes a local observable (e.g., the magnetization).  
    (b) Calculation of the two-point correlator with CTMRG environments.
    (c) Observables can be computed with either $L$ or $R$ tensors and the environments must be equivalent for consistency of the results. (In practice, this equivalence is ensured by the condition shown in the figure. This condition holds to high accuracy for the converged CTM environments.)
    (d) Definition of the transfer matrix $E$.
    }
\end{figure}

The converged CTM environments enable computation of both local observables and nonlocal correlation functions. 
Figures~\subfigref{fig:CTMRG}{a} and \subfigref{fig:CTMRG}{b} show tensor contractions, which are used in calculations of the one- and two-site observables.  
$T_{\rm imp}$ in Fig.~\ref{fig:Observables} corresponds to the impurity tensor similar to the iPEPS double-layer site tensor with a spin operator. 
To obtain the expectation value of these one- and two-site observables, the tensor contractions in Fig.~\ref{fig:Observables} must be additionally normalized by the same contractions with tensors $T_{\rm imp}$ replaced by $T$. Note that the one-site observable is computed only with the $L$-type tensors. But from the definitions of $L$ and $R$ tensors it follows that these quantities can also be computed only with the $R$-type tensors (or with some combination of $L$ and $R$). For consistency, all these definitions of observables must agree. In practical calculations, this agreement is enforced by the equality schematically depicted in Fig.~\subfigref{fig:Observables}{c}. 
This equality holds to a high accuracy for the converged CTM tensors (the detailed accuracy slightly depends on the choice of parameters in the randomized SVD algorithm). 
Finally, in our iPEPS calculations we use extrapolation of the results based on the correlation length scaling. Figure~\subfigref{fig:Observables}{d} defines the transfer matrix~$E$ with the honeycomb-lattice CTM tensors. The correlation length $\xi$ can be computed from the eigenvalues $\lambda_{i}$ of the transfer matrix $E$ as $\xi = - 
{1}/{\log{|\lambda_{2}/\lambda_{1}|}}$.

\subsection{Heisenberg model} 
Our variational ansatz \asr{for the Heisenberg model} is the iPEPS wave function consisting of uniform and $C_{3v}$-invariant tensors $A$ on all sites of the lattice. 
For the computation of observables, first, we map the tensor~$A$ onto the double-layer tensor~$T$ and then compute the corresponding environments with the above-specified prescription. 
Since the tensor is invariant under rotations and translations, for the energy computation we need only to compute  the single-bond interaction energy, which is mapped onto the two-site correlation function as described above. 
We compute this correlation function using only differentiable operations and then apply backwards differentiation to obtain the energy gradients in the space spanned by the original tensor $A$. 
These gradients serve for optimization of the iPEPS tensor $A$ using the limited-memory BFGS (L-BFGS) method. The implementation also relies on a set of \textsc{julia} numerical packages \textsc{zygote}, \textsc{backwardslinalg}, and \textsc{optim} \cite{Zygote.jl-2018,mogensen2018optim}.

The gradients are computed using the backward differentiation through the full CTM iteration procedure. This is in contrast with the more involved but less memory-consuming differentiation of the fixed-point equations proposed in Ref.~\cite{Liao_2019}. The CTM iterations consist only of tensor contractions and SVD or eigenvalue decompositions. For the tensor contractions, gradients can be defined as for usual multiplications and summations, while for the SVD and eigenvalue decompositions the gradients can be found in Ref.~\cite{giles2008extended} (gradients for iterative eigensolvers were derived more recently in Ref.~\cite{PhysRevB.101.245139}). Note that here the gradients are additionally symmetrized to prevent the loss of $C_{3v}$ invariance of the iPEPS tensor $A$.

We focus on the antiferromagnetic Heisenberg model on the honeycomb lattice with the Hamiltonian
\begin{equation}
    h = \sum_{\langle ij \rangle} S_{i}^{x} S_{j}^{x}+S_{i}^{y} S_{j}^{y}+S_{i}^{z} S_{j}^{z},
\end{equation}
where $\langle ij \rangle$ denotes summation over the nearest-neighbor sites $i$ and $j$. 
This model was already studied by means of the simple update iPEPS~\cite{PhysRevLett.101.090603,zhao2010renormalization}, quantum Monte Carlo simulations~\cite{low2009properties, jiang2012high, Reger_1989, PhysRevB.73.054422}, Schwinger boson mean-field theory~\cite{PhysRevB.49.3997}, series expansion~\cite{PhysRevB.45.9834}, spin-wave analysis~\cite{PhysRevB.44.11869}, and coupled cluster method~\cite{PhysRevB.89.184407}.

Due to the bipartite structure of the honeycomb lattice, the ground state is characterized by the antiferromagnetic order with a two-site unit cell. 
The model can be mapped into the model with the single-site unit cell by acting with a unitary transformation $\sigma_{y}$ on all sites of the one sublattice. Below, we also restrict ourselves to the real-valued tensors, but do not use an additional U(1) symmetry. The generalization to U(1)-symmetric iPEPS can be conducted along the lines of Ref.~\cite{hasik2021investigation}, where U(1) charges were found from the unrestricted optimization.

We perform the iPEPS optimization for the bond dimensions $D\in[2,7]$ and the CTMRG-environment dimensions $\chi \in [20, 120]$. 
After that, we calculate the observables in the framework of the honeycomb-lattice CTMRG with the increased dimension~$\chi$ of the environment. In particular, we increase $\chi$ up to $200$ for the energy and magnetization and up to $300$ in the analysis of the correlation length $\xi$.  
The latter exhibits very slow convergence to its infinite-$\chi$ value. 
For the extrapolation of $\xi$ to this limit, we use the scaling formula~\cite{Lauchli, PhysRevX.8.041033}
\begin{equation}
    \frac{1}{\xi(\chi)} = \frac{1}{\xi(\infty)} + a \log{\left(\frac{\lambda_{2}(\chi)}{\lambda_{4}(\chi)}\right)},
\end{equation}
where $\lambda_{2}$ and $\lambda_{4}$ are the second and the fourth largest eigenvalues of the transfer matrix $E$, respectively \footnote{Note that the original formula contains the second and the third largest eigenvalues, but in the Heisenberg model the second and the third transfer matrix eigenvalues are degenerate due to the U(1) symmetry.}.

\begin{figure}[t]
  \includegraphics[width=\linewidth]{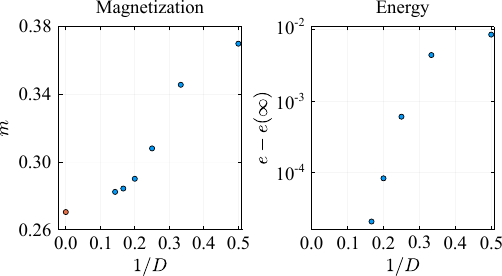}
  \caption{\label{fig:results}%
     Results of the variational calculations with iPEPS for the antiferromagnetic Heisenberg model. 
     The energy is plotted with respect to the extrapolated $\xi\to\infty$ value
     and demonstrates fast convergence. The staggered magnetization converges significantly slower, which is typical for the iPEPS calculations of gapless systems. 
    }
\end{figure}

The two main observables are the energy per site $e=\langle h\rangle/N$ and the staggered magnetization $m\equiv|m_i| = \sqrt{\langle S_{i}^{x} \rangle ^{2} +\langle S_{i}^{y} \rangle ^{2}+\langle S_{i}^{z} \rangle ^{2}}$.
In Fig.~\ref{fig:results} we show results for these observables depending on the bond dimension $D$ of the optimized iPEPS. 
To extrapolate the results to the infinite-$D$ limit, we use the dependencies of the magnetization~$m$ and the energy~$e$ on the correlation length~$\chi$ determined in Ref.~\cite{Lauchli}:
\begin{align}
    m^{2}(\xi) &= m^{2}(\infty) + \frac{a}{\xi}, \label{eq:magn}
    \\
    e(\xi) &= e(\infty) + \frac{b}{\xi^{3}}. \label{eq:energy}
\end{align}

\begin{figure}[t]
  \includegraphics[width=\linewidth]{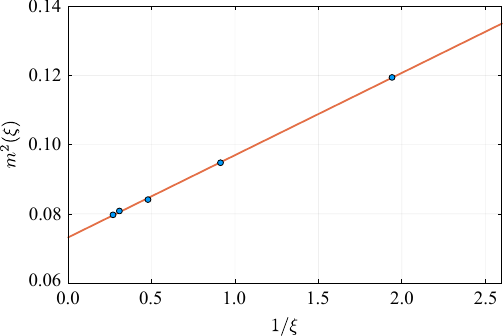}
  \caption{\label{fig:magnetisation}%
        Dependence of the square of magnetization $m^{2}(\xi)$ for $D \in [3,7]$ on the inverse of the correlation length $\xi$ and the result of the linear fit. 
    }
\end{figure}

The dependence of the staggered magnetization on the correlation length $\xi$ is shown in Fig.~\ref{fig:magnetisation}. 
It clearly shows a linear behavior in accordance with the scaling formula~\eqref{eq:magn}. 
The linear fit yields the value of the magnetization $m(\infty) = 0.2705(25)$, which can be compared to $m = 0.27$ from the series expansion~\cite{PhysRevB.45.9834}, $m = 0.24$ from the spin-wave analysis~\cite{PhysRevB.44.11869}, and to the (seemingly) most accurate Monte Carlo (MC) result $m_{\rm MC} = 0.268\,82(3)$~\cite{jiang2012high}. The error was estimated from both the error of the least-squares fit and from partial fits with the reduced number of points. Note that the values of magnetization, obtained directly from the optimized iPEPS wave functions, are noticeably higher ($m > 0.28$) that partially explains the higher error.

\begin{figure}[t]
  \includegraphics[width=\linewidth]{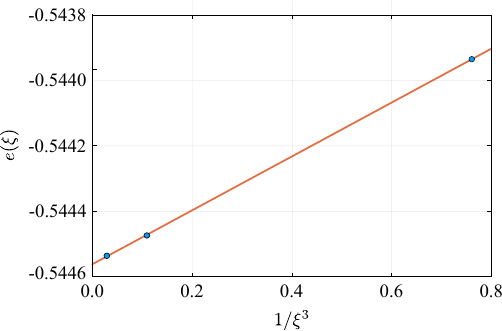}
  \caption{\label{fig:energy}%
        Dependence of the energy per site $e(\xi)$ for $D \in [4,6]$ on the the third power of the inverse of the correlation length $\frac{1}{\xi^{3}}$ and the result of the linear fit. 
    }
\end{figure}

In Fig.~\ref{fig:energy} we show the dependence of the energy per site $e$ on ${1}/{\xi^{3}}$ for $D \in [4,6]$, which also demonstrates a clear linear behavior in agreement with Eq.~\eqref{eq:energy}. 
At smaller $D$ (not shown in the figure), there are deviations from the ${1}/{\xi^{3}}$ dependence. For $D=6$ we obtain the energy $e = -0.544\,536$, while the energy extrapolation using the scaling formula~\eqref{eq:energy} yields $e(\infty) = -0.544\,563(11)$. The quantum Monte Carlo prediction is $e_{\rm MC} = -0.544\,553(20)$~\cite{low2009properties}, hence, both the energy~$e(\xi)$ for $D=6$ and the extrapolated~$e(\infty)$ are within the MC error bars.

\subsection{Kitaev model}
\begin{figure}[t]
  \includegraphics[width=\linewidth]{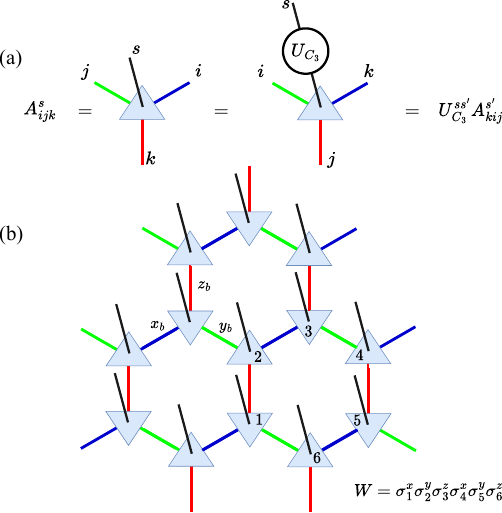}
  \caption{\label{fig:Kitaev_illustration}%
        (a) The symmetry properties of the tensor~$A$ representing the iPEPS wave function in the Kitaev model. (b) Spatial arrangement of the bonds $x_b$, $y_b$, and $z_b$ in the Kitaev model and the definition of the flux $W$.
    }
\end{figure}

Our CTMRG approach can not be directly applied to Hamiltonians with more general anisotropic structure of couplings, since it requires a more involved truncation procedure. We leave the development of this truncation procedure (and also the generalization to the larger unit cells) to future research. Still, there exists an important class of anisotropic Hamiltonians, which can be directly simulated with the generalization of our method. These are Hamiltonians of the Kitaev model and the corresponding generalizations known as the Kitaev-Heisenberg and $K$-$\Gamma$ models. The ferromagnetic Kitaev model has the following Hamiltonian, 
\begin{equation}\label{Ham_Kitaev}
    H = -J\sum_{x_b,ij}
    S^{x}_{i} S^{x}_{j} - J\sum_{y_b,ij}
    S^{y}_{i} S^{y}_{j} - J\sum_{z_b,ij}
    S^{z}_{i} S^{z}_{j},
\end{equation}
where the sums are taken over certain types of bonds $r_b=\{x_b,y_b,z_b\}$ on the lattice (see also Fig.~\subfigref{fig:Kitaev_illustration}{b}). We hereafter fix $J=1$. The model is exactly solvable by the fermionization procedure~\cite{KITAEV20062}. Its ground state at these values of parameters is a gapless spin liquid with the vanishing magnetization and the integral of motion $W=1$, which is schematically shown in Fig.~\subfigref{fig:Kitaev_illustration}{b}.

From the Hamiltonian~\eqref{Ham_Kitaev} it is clear that the model is not invariant under regular rotations and reflections. However, it remains invariant under transformations that belong to the modified symmetry group consisting of rotations and reflections followed by the unitary transformations on the physical Hilbert space $U_{C_{3}} R$ and $U_{p,i} p_{i}$, where $R$ is the lattice rotation, and $p_{i}$ is the lattice reflection with respect to the axis $i$. $U_{C_{3}}$ and $U_{p,i}$ are the unitary matrices, which act on the physical Hilbert space in the way to complete lattice rotations and reflections into the model symmetries. These matrices are defined as
\begin{align}
    U_{C_{3}} = - \exp{\left[i\frac{2\pi}{3}\right]}(I + i\sigma_{x} + i\sigma_{y} +i \sigma_{z})/2, \\
    U_{p,z} = -(\sigma_{x} - \sigma_{y})/\sqrt{2},
\end{align}
where $I$ and $\sigma_i$ are the standard $2\times2$ identity and Pauli matrices, respectively.
The matrices $U_{p,y} = -\exp{\left[i\frac{2\pi}{3}\right]}(\sigma_{z} - \sigma_{x})/\sqrt{2}$ and $U_{p,x} = -\exp{\left[-i\frac{2\pi}{3}\right]}(\sigma_{y} - \sigma_{z})/\sqrt{2}$ are defined analogously.
The model is additionally invariant under the time reversal, $\tau = i\sigma_{y} K$, where $K$ is the complex conjugation. 
It is then natural to require the iPEPS tensor $A$ to be invariant under the modified $C_{v3}$ symmetry consisting of the simultaneous rotation with the unitary transformation and  reflection-conjugation modified to sequential application of the modified reflection and time reversal $\tau$. 
This rotation and rotation-conjugation symmetry fixes the magnetization of iPEPS to be oriented along the direction $(1,1,1)$ in the spin space. 
Note that tensors with the same symmetry properties were also used in Ref.~\cite{LGOperator}. 
The symmetric tensor can be obtained by the application of a projector $P$ of the form $P = I + U_{C_{3}} R + U^{2}_{C_{3}} R^{2} + i\sigma^{y} K(U_{p,x} p_{x}+U_{p,y} p_{y} + U_{p,z} p_{z})$ on the initially arbitrary iPEPS tensor $A^{s}_{ijk}$, where $U_{C_{3}} R A^{s}_{ijk} = U_{C_{3}}^ {ss'} A^{s'}_{kij}$ and $U_{p,x} p_{x} A^{s}_{ijk} = U_{p,x}^{ss'} A^{s}_{ikj} $. The action of this symmetry transformation is shown in Fig.~\subfigref{fig:Kitaev_illustration}{a}.

Next, we should mention the following property of the double-layer tensor $T = \sum_{s} A^{s}_{ijk} A^{\dagger,s}_{i'j'k'}$, which enters the CTM transform: It is invariant under rotations and reflection-conjugations due to cancellation of the unitary matrices $U_{C_{3}}$ and $U_{\sigma,i}$ in the double-layer contraction, thus one can apply the  above-specified CTMRG approach to the given iPEPS ansatz, even if the iPEPS tensors $A$ are not rotationally invariant and the model Hamiltonian is anisotropic. 

\begin{figure}[t]
  \includegraphics[width=\linewidth]{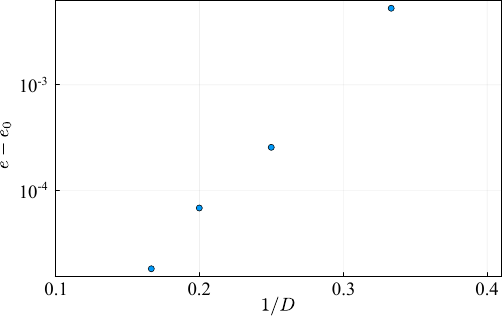}
  \caption{\label{fig:energy_Kitaev}%
        Dependence of the energy per site $e$ of the Kitaev model on the inverse of the bond dimension $D$. The energy is plotted with respect to the exact value $e_0 = -0.196\,82$. 
    }
\end{figure}

The ground-state energy of the Kitaev model~\eqref{Ham_Kitaev} is determined with the exact diagonalization in the original paper~\cite{KITAEV20062}, where the energy per site is equal to $e_0 = -0.196\,82$.  
Our results for the energy per site for different values of the bond dimension~$D$ are shown in Fig.~\ref{fig:energy_Kitaev}. 
In particular, at $D=6$ the energy per site is equal to $e = -0.196\,807$. Other observables are also comparable to the exact diagonalization predictions: For $D \geq 4$ the magnetization $m < 3\times10^{-4}$ and $|1-W| < 1.5\times10^{-4}$. 
The correlation length~$\xi$ becomes extremely large, in particular, $\xi = 368$ at $D=4$, meaning that the iPEPS wave function is nearly critical. Note that our results at $D=4$ are only marginally better than the ones from Ref.~\cite{LGOperator}, which confirms the efficiency of their loop gas ansatz. 

For more general anisotropic Hamiltonians, it is typically impossible to use the iPEPS ansatz with the rotationally invariant two-layer tensor $T$. 
Therefore,  a more general CTMRG scheme must be introduced for the anisotropic tensor~$T$, which should include several corner matrices $C$ for different directions and a new iteration procedure. The same procedure should be introduced to deal with the enlarged unit cells.  It looks promising to develop the modified CTMRG scheme for these cases along the lines of Ref.~\cite{PhysRevB.82.245119}.

\section{Conclusion and Outlook} 
In this paper we have  realized the iPEPS variational optimization within the automatic differentiation of the honeycomb-lattice corner transfer matrix renormalization group. We tested the method on the corresponding antiferromagnetic Heisenberg and ferromagnetic Kitaev models and obtained the results comparable 
to the ones from the state-of-the-art quantum Monte Carlo simulations and the exact diagonalization approach.

This work opens several future research directions. The algorithm can be naturally applied to the frustrated Heisenberg antiferromagnets with NNN and even longer-range interactions~\cite{FrustratedAF1, FrustratedAF2, FrustratedAF3, FrustratedAF4, FrustratedAF5, FrustratedAF6, FrustratedAF7}. Upon certain technical modifications, it can also be employed for the star lattice geometry~\cite{Jahromi_2018_star}. 
Furthermore, it would be interesting to generalize the method to larger unit cells~\cite{Baxter1999}. Another possible research direction is the application of the variational iPEPS to hyperbolic lattice geometries~\cite{Dani_ka_2015, Dani_ka_2016}, which were recently realized in experiments~\cite{kollar2019hyperbolic,boettcher2020quantum}. 

Finally, the automatic differentiation can be effectively applied to TRG and its modifications~\cite{Liao_2019, chen2020automatic}. This opens a way for the variational optimization of iPEPS with the tensor renormalization. 

\acknowledgements
The authors acknowledge support from the National Research Foundation of Ukraine, Grant No.~0120U104963, the Ministry of Education and Science of Ukraine, Research Grant No.~0122U001575, and the National Academy of Sciences of Ukraine, Project No. 0121U108722.

\bibliography{HCTMRG}

\end{document}